\documentclass[aps,prl,longbibliography,twocolumn,superscriptaddress]{revtex4-2}

\usepackage[dvipdfmx]{graphicx}
\usepackage{txfonts}
\usepackage{bm}
\usepackage{color}
\usepackage{hyperref}
\usepackage{booktabs}
\usepackage{here}
\begin{document}
\title{Finite-Size Corrections to Defect Energetics along One-Dimensional Configuration Coordinate}
 
\author{Yu Kumagai}
\email[]{yukumagai@tohoku.ac.jp}
\affiliation{Institute for Materials Research, Tohoku University, 2-1-1 Katahira, Aoba-ku, Sendai, 980-8577, Japan}

\date{\today}

\definecolor{blue}{rgb}{0.0,0.0,1}
\hypersetup{colorlinks,breaklinks,
            linkcolor=blue,urlcolor=blue,
            anchorcolor=blue,citecolor=blue}

\begin{abstract}
Recently, effective one-dimensional configuration coordinate diagrams have been utilized to calculate the line shapes of luminescence spectra and non-radiative carrier capture coefficients via point defects. Their calculations necessitate accurate total energies as a function of configuration coordinates. Although supercells under periodic boundary conditions are commonly employed, the spurious cell size effects have not been previously considered. In this study, we have proposed a correction energy formalism and verified its effectiveness by applying it to a nitrogen vacancy in GaN. 
\end{abstract}

\maketitle

\begin{figure}
  \includegraphics[width=1\linewidth]{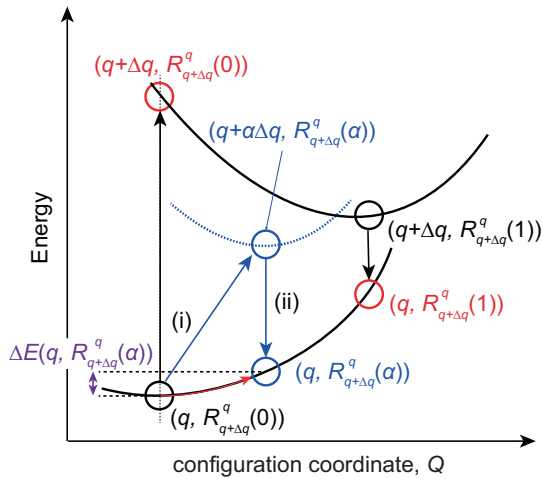}
  \caption{ Schematics of the configuration coordinate diagram between charge $q$ and $q+\Delta q$, where $\Delta q=\pm1$. 
  The black circles depict energies of structures relaxed at each charge, whereas the red circles represent those relaxed at different charges. 
  The blue circles denote the energy in virtual state $\left(q+\alpha\Delta q,\ R_{q+\Delta q}^q(\alpha)\right)$ (see text for details). 
  The correction energy for $\left(q,\ R_{q+\Delta q}^q(\alpha)\right)$ is equal to the sum of correction energies for route (i) and (ii).}
  \label{ccd_schematic}
\end{figure}

$Introduction.-$ 
Point defects are ubiquitous in solids and play important roles in many fundamental physics (e.g., superconductors~\cite{Bednorz.1986} and topological materials~\cite{Ando.2013}), and practical applications (e.g., solar cell absorbers~\cite{Nayak.2019} and light-emitting diodes~\cite{Nakamura.1992}). 
To better understand and even predict the point defect properties, first-principles calculations have been routinely used over the last few decades~\cite{RevModPhys.86.253}. 
Recently, the line shapes of luminescence spectra~\cite{Alkauskas.2012,Frodason.2017} and carrier capture coefficients~\cite{Alkauskas.2014,Kim.2019,Zhang.2020,Kavanagh.2022} via point defects have been calculated from first principles. The key approximation in these calculations is to simplify the defect-related phonon modes to one effective mode~\cite{Alkauskas.2014} that is parallel to the distortion of the defect geometry when gaining or losing an electron. Total energies as a function of such a one-dimensional (1D) configuration are schematically described with a configuration coordinate (CC) diagram as shown in Fig.~\ref{ccd_schematic}.

To accurately determine the formation energies of charged defects using supercells under periodic boundary conditions (PBCs), it is necessary to account for finite cell size errors that result from spurious electrostatic interactions. Formalisms have been developed for making such corrections, and the Freysoldt-Neugebauer-Van de Walle (FNV) method~\cite{Freysoldt.2009} is currently recognized as the most advanced approach. 
We have expanded the FNV method to accommodate anisotropic materials with atomic relaxation (eFNV)~\cite{Kumagai.2014i}. 
Optical transition levels via point defects, also known as vertical transition (VT) energies, can be computed assuming that the atomic configuration remains unchanged during the electron transitions, in accordance with the Franck-Condon principle~\cite{Franck.1926,Condon.1926}.
Their calculations require a distinct correction methodology, for which we have proposed a method termed the GKFO method~\cite{Gake.2020}. 
Falletta et al. have also reported a comparable correction energy from a different viewpoint~\cite{Falletta.2020}.

So far, when calculating the CC diagram, all the energies along the CC were constantly shifted by the correction energies obtained at the ground state using the FNV method.
Unfortunately, this approach is incorrect. 
 To understand why, let us begin by defining some notation.

 The atomic configuration $R$ at the displaced configuration from the relaxed structure at charge $q$ to the relaxed structure at charge $q$+$\Delta q$ ($\Delta q=\pm1$) in a ratio of $\alpha$ is denoted as  $R_{q+\Delta q}^q(\alpha)$. 
Note that $\alpha$ may take a value other than 0 to 1. 
The relaxed atomic configuration at charge $q$ is denoted as $R_{q+\Delta q}^q(0)$, and $R_{q+\Delta q}^q(\alpha)= R_q^{q+\Delta q}(1-\alpha)$ is always satisfied. 
The defect state at charge $q$ and configuration $R_{q+\Delta q}^q(\alpha)$ and its energy are denoted as $\left(q, R_{q+\Delta q}^q(\alpha)\right)$, and $E\left(q,\ R_{q+\Delta q}^q(\alpha)\right)$, respectively. 
We also defined its relative energy with respect to $E\left(q, R_{q+\Delta q}^q(0)\right)$ as $\Delta E\left(q,\ R_{q+\Delta q}^q(\alpha)\right)$ (see Fig.~\ref{ccd_schematic}). 

In the case of $q=0$, $E\left(0, R_{\Delta q}^0(0)\right)$ does not require any electrostatic correction. 
However, when e.g. $\alpha=1$, the neighboring ions move slightly as if charge $\Delta q$ is located at the defect site. 
Such ion distortion occurs linearly with respect to $\alpha$ and is effectively regarded as a polarization charge~\cite{Falletta.2020}, which interacts with its images under PBCs and requires energy corrections as a function of the CC. 
Our objective is to develop a formalism for correcting these energies and verify its effectiveness.

To obtain the correction energy for $\Delta E\left(q,\ R_{q+\Delta q}^q(\alpha)\right)$ ($\Delta E_{\rm cor}$), we introduce a virtual state $\left(q+\alpha\Delta q,\ R_{q+\Delta q}^q(\alpha)\right)$, in which the ionic dipoles is exactly balanced with the defect charge $q+\alpha\Delta q$.
While computing such a state explicitly is computationally demanding, it is not necessary in the final formulation.
$\Delta E_{\rm cor}$ is the sum of the (i) eFNV and (ii) GKFO correction energies, as illustrated in Fig~\ref{ccd_schematic}. 
Initially, we consider the cubic system and later generalize it. 

The correction energies for the relaxed structures with charge $q$ and $q+\alpha\Delta q$ are expressed in the eFNV approach as $E_{\mathrm{PC}}^q\left(\varepsilon_0\right)-qC^q$ and  $E_{\mathrm{PC}}^{q+\alpha\Delta q}\left(\varepsilon_0\right)-\left(q+\alpha\Delta q\right)C^{q+\alpha\Delta q}$, respectively. 
Here, $E_{\mathrm{PC}}^q\left(\varepsilon\right)$ is the point-charge (PC) correction energy for charge $q$ screened by a dielectric constant $\varepsilon$ and $C^q$ and $C^{q+\alpha \Delta q}$ are alignment constants chosen such that the short-range potential decays to zero far from the defect (see Refs.~\cite{Freysoldt.2009,Kumagai.2014i}). 
$\varepsilon_0$ is the static dielectric constant, which is the sum of the ion-clamped ($\varepsilon_\infty$) and ionic ($\varepsilon_{\rm ion}$) dielectric constants.

The correction energy for route (i) is then described as 
\begin{eqnarray}
{\Delta E}_{\mathrm{cor}}^{\mathrm{(i)}}&=&E_{\mathrm{PC}}^{q+\alpha\Delta q}\left(\varepsilon_0\right) \nonumber \\
  &-&\ E_{\rm PC}^q\left(\varepsilon_0\right)-\left[\left(q+\alpha\Delta q\right)C^{q+\alpha\Delta q}-qC^q\right].
\label{eq1}
\end{eqnarray}

Route (ii) corresponds to the VT, where the charge transitions from $q+\alpha\Delta q$ to $q$, and it is corrected with the GKFO method as follows:
\begin{eqnarray}
\Delta E_{\mathrm{cor}}^{\mathrm{(ii)}}=-\frac{2\alpha\Delta q}{q+\alpha\Delta q}\ E_{\rm PC}^{q+\alpha\Delta q}\left(\varepsilon_0\right)+E_{\mathrm{PC}}^{-\alpha\Delta q}\left(\varepsilon_\infty\right) \nonumber \\
- \left(-\alpha\Delta q C^{-\alpha\Delta q}-\alpha\Delta q C^{q+\alpha\Delta q}+\frac{\varepsilon_\infty}{\varepsilon_0}\left(q+\alpha\Delta q\right)C^{-\alpha\Delta q}\right).
\label{eq2}
\end{eqnarray}
Here, $C^{-\alpha\Delta q}$ is an alignment constant caused by introducing additional charge $-\alpha\Delta q$ without altering the atomic configuration $R_{q+\Delta q}^q(\alpha)$~\cite{Gake.2020}.
As derived by Komsa et al.~\cite{Komsa.2012g9}, it can be written as:
\begin{eqnarray}
C^{-\alpha\Delta q}=\frac{2\pi \int{d {\bm r}^3\Delta\rho^{-\alpha\Delta q}r^2}}{\varepsilon_\infty L^3},
\label{eq3}
\end{eqnarray}
where $L$ is the side length of the supercell, and $\Delta\rho^{-\alpha\Delta q}$ is the defect charge distribution modified by adding charge $-\alpha\Delta q$.

The sum of the PC correction terms can be expressed as 
\begin{eqnarray}
{\Delta E}_{\mathrm{cor}}^{\mathrm{PC}}\ =\ E_{\mathrm{PC}}^{-\alpha\Delta q}\left(\varepsilon_\infty\right)-E_{\mathrm{PC}}^{-\alpha\Delta q}\left(\varepsilon_0\right) 
\label{eq4}
\end{eqnarray}
(see Supplemental Materials for the detailed derivation). 
Because $\Delta q=\pm1$ and the sign of $\Delta q$ is irrelevant for $E_{\mathrm{PC}}^{-\alpha\Delta q}$, $\Delta E_{\mathrm{cor}}^{\mathrm{PC}}$ depends on neither initial charge $q$ nor $\Delta q$ but on $\alpha$. 
Note, however, that $\Delta E_{\rm cor}$ represents the additional correction to the $relative$ energy with respect to the relaxed structure, and the total correction energy is the sum of $\Delta E_{\rm cor}$ and the FNV correction energy to $E\left(q, R_{q+\Delta q}^q(0)\right)$, where the PC term depends on $q^2$.

The summation of the alignment terms is written as 
\begin{eqnarray}
\Delta E_{\mathrm{cor}}^{\mathrm{align}} = -\left(q+\alpha\Delta q\right)C^{q+\alpha\Delta q}+qC^q \nonumber \\
 + \left(\alpha\Delta q C^{-\alpha\Delta q}+\alpha\Delta q C^{q+\alpha\Delta q}-\frac{\varepsilon_\infty}{\varepsilon_0}\left(q+\alpha\Delta q\right)C^{-\alpha\Delta q}\right).   
\label{eq5}
\end{eqnarray}
We define $C^{\rm align}$ as the alignment term estimated between $\left(q,\ R_{q+\Delta q}^q(\alpha)\right)$ and $\left(q,\ R_{q+\Delta q}^q(0)\right)$ by removing the long-range potential of charge $-\alpha\Delta q$ screened by $\varepsilon_{\rm eff}$, where $1/\varepsilon_{\rm eff}=1/\varepsilon_\infty-1/\varepsilon_0$, and can be written as 
\begin{eqnarray}
C^{\rm align}=\frac{2\pi\int{d\bm{r}^\mathbf{3}\Delta\rho^{-\alpha\Delta q}r^2}}{\varepsilon_{\rm eff}L^3}   
\label{eq6}
\end{eqnarray}
as in Eq.~\ref{eq3}. 
Using $C^{\rm align}$, we obtain $\Delta E_{\mathrm{cor}}^{\mathrm{align}}=\alpha\Delta qC^{\rm align}$ (see Supplemental Materials for details of the derivation). 
Thus, the $\Delta E_{\rm cor}$ is expressed as 
\begin{eqnarray}
{\Delta E}_{\rm cor}\ =\ E_{\mathrm{PC}}^{-\alpha\Delta q}\left(\varepsilon_{\rm eff}\right)-\left(-\alpha\Delta q\right)C^{\rm align}. 
\label{eq7}
\end{eqnarray}
The obtained ${\Delta E}_{\rm cor}$ is simply estimated by considering the effective charge $-\alpha\Delta q$ screened by an effective dielectric constant $\varepsilon_{\rm eff}$. This can be easily extended to the anisotropic systems by replacing $\varepsilon_{\rm eff}$ with the tensor form, ${\bar{\varepsilon}}_{\rm eff}$. 
Furthermore, it is applicable to the anharmonic potential energy surface~\cite{Kim.2019} because the ionic displacement distance for the screening linearly depends on $\alpha$.

\begin{figure*}
  \includegraphics[width=1\linewidth]{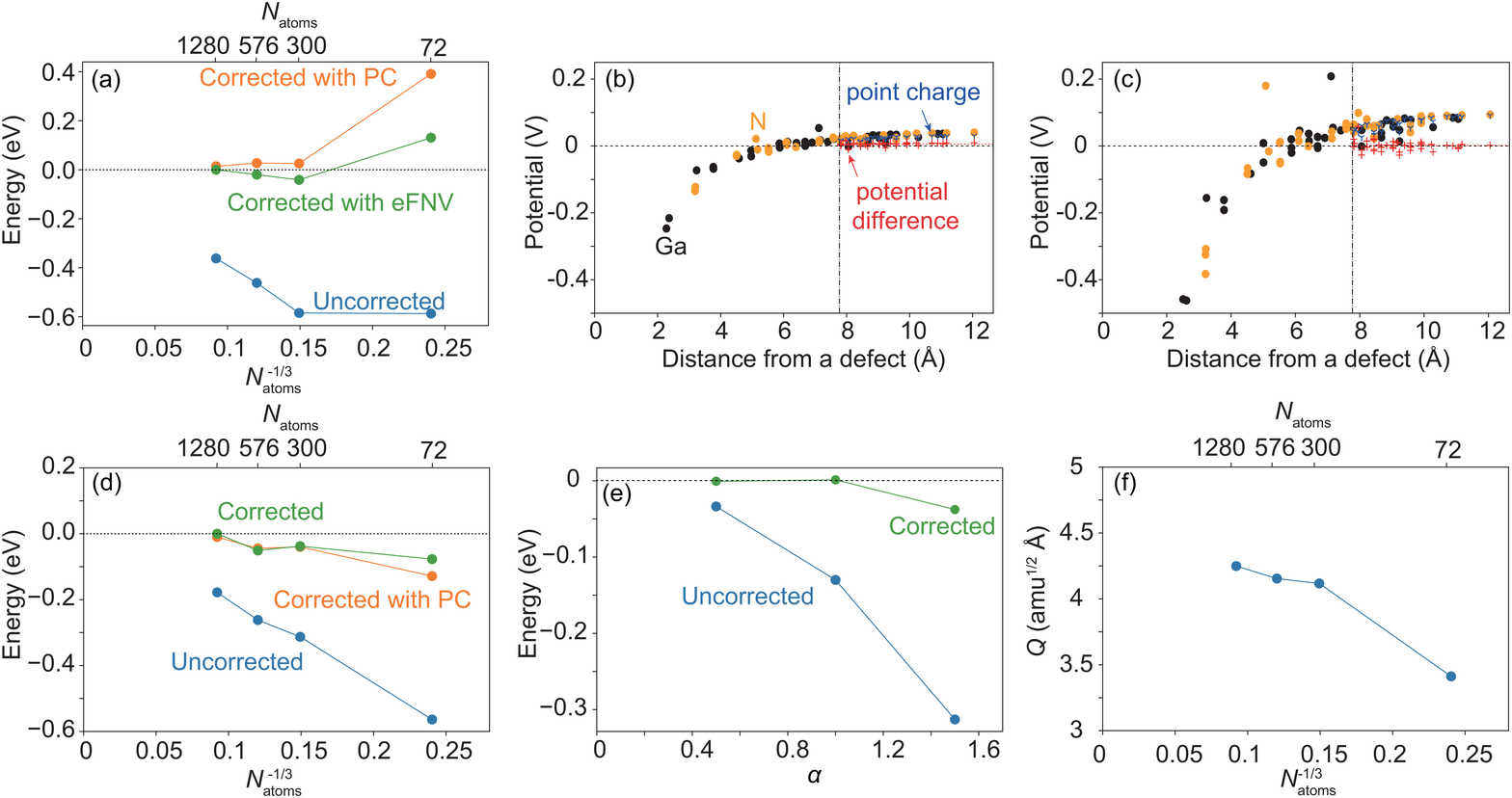}
  \caption{
  (a) Relative formation energies of $V_{\rm N}^{+2}$ in GaN without any corrections, with the PC corrections, and with the eFNV corrections as a function of the supercell size.  
  The zero energy value is set to the energy corrected with the eFNV method at the largest supercell.
  (b, c) Atomic site potentials caused by $V_{\rm N}^{+2}$ at (a) $\left(2,\ R_3^2(0.5)\right)$ and (b) $\left(2,\ R_3^2(1.5)\right)$ in the 576-atom supercell that are calculated by subtracting the potentials in the pristine supercell.
  Blue marks are the electrostatic potentials caused by periodic point charges $-\alpha\Delta q$ and background charge screened by an effective dielectric tensor, while red crosses are their differences from the atomic site potentials. 
  The vertical dashed lines indicate the radius of the sphere in contact with the supercell.
  The averaged potential difference far from the defect ($C^{\rm align}$) is evaluated outside of the sphere (see Ref.~\cite{Kumagai.2014i} for details). 
  (d) The supercell size dependence of $\Delta E\left(2,\ R_3^2(1.5)\right)$ without any corrections, with PC corrections, and with corrections in Eq.~\ref{eq7}. 
  (e) $E\left(2,\ R_3^2(\alpha)\right)$ as a function of $\alpha$ without and with corrections in Eq.~\ref{eq7}. 
  (f) The configuration coordinate $Q$ of $V_{\rm N}^{+2}$ in GaN at $R_3^2(1)$. 
In (d) and (e), the zero energy values are set  to the energies corrected with Eq.~\ref{eq4} at the largest supercells.}
  \label{correction_results}
\end{figure*}

To verify the effectiveness of the correction formalism in Eq.~\ref{eq7}, we have calculated the total energies of the nitrogen vacancy ($V_{\rm N}$) in GaN. 
To demonstrate that the initial charge state does not affect $\Delta E_{\rm cor}$, we calculated the energies at $q=+2$ along the CC to the structure at $q=+3$.
First, we confirm that the conventional eFNV method works well for calculating the formation energy of $V_{\rm N}^{+2}$ ($E(V_{\rm N}^{+2}))$ in GaN as shown in Fig.~\ref{correction_results}. Specifically, $E(V_{\rm N}^{+2})$ is well corrected when using the 300-atom supercell. 

In Figs.~\ref{correction_results}(b) and (c), we present the atomic site potentials caused by $V_{\rm N}^{+2}$ at two different configuration coordinates, namely $\left(2,\ R_3^2(0.5)\right)$ and $\left(2,\ R_3^2(1.5)\right)$, in the 576-atom supercell. 
Despite having identical charge $q=+2$, a significant difference in the potential variation is observed between the two coordinates. 
The atomic site potentials calculated from a charge $-\alpha\Delta q$ at the defect site screened by ${\bar{\varepsilon}}_{\rm eff}$ are also displayed. 
These potentials adequately reproduce tendencies of the atomic site potentials in the region far from the defect, leading to an estimation of $C^{\rm align}$.

Figure~\ref{correction_results}(d) presents $\Delta E\left(2,\ R_3^2(1.5)\right)$ of $V_{\rm N}^{+2}$ as a function of number of atoms in the pristine supercell ($N_{\rm atoms}$). 
When applying the constant energy correction irrelevant to the CC that corresponds to “Uncorrected” in this Letter, there are noticeable finite-cell size errors. 
These errors are effectively eliminated when $\Delta E_{\rm cor}$ in Eq.~\ref{eq7} are applied, thereby validating our formalism. 
As shown in Fig.~\ref{correction_results}(d), the alignment terms do not exert a significant effect, meaning that the ionic dipoles that screen the additional defect charge $\Delta q$ do not appreciably alter the defect charge.

Figure~\ref{correction_results}(e) depicts $\Delta E\left(2,\ R_3^2(\alpha)\right)$ with and without corrections, as a function of the displacement fraction $\alpha$. 
As Eq.~\ref{eq7} suggests, the finite size errors are predominantly dependent on  $\alpha^2$. 
When computing the non-radiative carrier capture coefficients, the position at which the potential energy surfaces of two states intersect has a significant effect on the capture rate. 
Such intersections can occur at values of  $\alpha$ greater than 2, as exemplified by the carbon impurity at the nitrogen site (${\rm C}_{\rm N}$) in GaN~\cite{Alkauskas.2014}. 
Hence, it is likely that the intersection coordinates and energies are significantly modified by applying $\Delta E_{\rm cor}$. 
It is also worth noting that a miscalculation of 0.1 eV in the energies along the CC can result in a tenfold error in the non-radiative carrier capture coefficients~\cite{Alkauskas.2014}. 

As shown in Figs.~\ref{correction_results}(a) and (d), $E(V_{\rm N}^{+2})$ and  $\Delta E\left(q,\ R_{q+\Delta q}^q(\alpha)\right)$ are effectively corrected even using the 72-atom supercell. 
However, the difference of the configuration coordinate $Q$ (see Method) between charge $q$ and $q+\Delta q$ ($\Delta Q$) is not accurately estimated using the 72-atom supercell as shown in Fig.~\ref{correction_results}(f).
This might also be related to the fact that the 72-atom supercell calculation of $E(V_{\rm N}^{+2})$ shows a different tendency in Fig.~\ref{correction_results}(a).
Because errors in $Q$ have a significant impact on the calculations of the line shapes of luminescence spectra and carrier capture coefficients, we generally recommend using larger supercells, particularly to improve $Q$.

The effective dielectric constant $\varepsilon_{\rm eff}$ quantifies the polarization resulting from ion displacement.
It is related to the long-range electron-phonon coupling, and is commonly used when discussing self-trapped polarons~\cite{Emin.2013}. 
In general, decreasing $\varepsilon_{\rm eff}$ leads to an increase in $E_{\rm cor}$. 
Because $\varepsilon_{\rm eff}=\varepsilon_\infty+\ \varepsilon_\infty^2\ /\varepsilon_{\rm ion}$, $\varepsilon_{\rm eff}$ decreases as $\varepsilon_\infty$ decreases and $\varepsilon_{\rm ion}$ increases. 
To assess the potential magnitude of the error in realistic materials resulting from neglecting $\Delta E_{\rm cor}$ in Eq.~\ref{eq7}, we calculated $\varepsilon_{\rm eff}$ for 931 oxide materials for which we calculated the oxygen vacancies in our previous work~\cite{Kumagai.2021}, using the spherical averages of $\varepsilon_\infty$ and $\varepsilon_{\rm ion}$. 
The smallest $\varepsilon_{\rm eff}$, which is 2.5, is found in CsLi$_5$(BO$_3$)$_2$, where the averaged  $\varepsilon_\infty$ and $\varepsilon_{\rm ion}$ are 2.1 and 10.8, respectively.  
Then, $\Delta E_{\rm cor}$ for the oxygen vacancy at state $\left(1,\ R_2^1(\alpha=2)\right)$ is estimated to be 1.2 eV even using the very large 448-atom supercell.

We can estimate how much the energy correction modifies the effective phonon frequency, which is given by the equation $\Omega^2\ =\frac{\partial^2E}{\partial Q^2}$ in the harmonic approximation~\cite{Alkauskas.2012,Alkauskas.2014}.
Replacing $\Omega$ with $\Omega+\Omega_{\rm cor}$, where $\Omega_{\rm cor}$ is the corrected frequency, and $E$ with $E+\Delta E_{\rm cor}$, and assuming that $\Omega_{\rm cor} \ll \Omega$ and the PC term is dominant in Eq.~\ref{eq7}, we obtain:
\begin{eqnarray}
\Omega_{\rm cor}\ =\frac{E_{\mathrm{PC}}^{\Delta q}\left(\varepsilon_{\rm eff}\right)}{\Omega\left(\Delta Q\right)^2},  
\label{eq8}
\end{eqnarray}
(see Supplemental Materials for details of the derivation). 
For the case of $\left(2,\ R_3^2(\alpha)\right)$ for $V_{\rm N}$, we find ${\hbar\Omega}_{\rm cor} = 3.28$ meV when using the 72-atom supercell, which is 14.5 \% of the uncorrected $\hbar\Omega$. 
Note that, because ${\hbar\Omega}_{\rm cor}$ is approximately proportional to $E_{\mathrm{PC}}^{\Delta q}\left(\varepsilon_{\rm eff}\right)$, its convergence as a function of $N_{\rm atom}$ is slow if $\Delta E_{\rm cor}$ is not applied.

To conclude, we derived the correction formalism for the defect energies along the 1D CC, which has not been considered so far. 
The correction energy is described with the effective charge $-\alpha\Delta q$ and effective dielectric constant $\varepsilon_{\rm eff}$. 
Its effectiveness has been verified with a nitrogen vacancy in GaN. 
Since our correction method is automatically applied with negligible computational cost compared to first-principles calculations, we believe that it would be routinely used in the future.

{\it Method.} 
All the calculations were performed using the projector augmented-wave (PAW) method~\cite{Blochl.1994,Kresse.1998} implemented in VASP~\cite{Kresse.1996}.
We adopted the HSE06 hybrid functional~\cite{Krukau.2006}. 
A fraction of Fock exchange parameter was set to 0.31 to reproduce the experimental band gap~\cite{Alkauskas.2014}. 
See the Supplemental Materials for the computational details. 
The harmonic phonon frequencies were calculated using Nonrad~\cite{Turiansky.2021px}. 
All the VASP input settings were generated with the VISE code (version 0.7.0)~\cite{vise}, while the processing related to defects was done with pydefect~\cite{pydefect}. 
The configuration coordinate $Q$ is defined as $\sqrt{\sum_{\alpha}{M_\alpha\cdot\Delta \bm{R}_\alpha^2}}$, where $M_\alpha$ and $\Delta \bm{R}_\alpha$ are the atomic mass and the displacement vector from the equilibrium position of atom $\alpha$, respectively~\cite{Alkauskas.2012,Alkauskas.2014}. 
The ion-clamped dielectric constants ($\varepsilon_\infty$) for 931 oxides were obtained from the long-range wavelength limit in the real part of dielectric functions calculated using the dielectric dependent hybrid functional~\cite{Skone.2014} (see Ref.~\cite{Kumagai.2023} for details). 
The ionic contributions ($\varepsilon_{\rm ion}$) were calculated using the density functional perturbation theory~\cite{Gonze.1997} (see Ref.~\cite{Kumagai.2021} for details).

{\it Acknowledgement.} 
Discussion with Soungmin Bae is sincerely appreciated.
This study was financially supported by PRESTO (JPMJPR16N4) from the Japan Science and Technology Agency, KAKENHI (Grant No. 22H01755), and the E-IMR project at IMR, Tohoku University.

\end{document}